\documentclass[twocolumn,floatfix,prb,showpacs]{revtex4}
\usepackage{graphicx}
\usepackage{amssymb}
\usepackage{amsmath}
\usepackage{color}
\usepackage{psfrag}
\usepackage{epsfig}
\usepackage{bbm}
\usepackage{bm}
\newcommand{\ket}[1]{\left| #1 \right\rangle}
\newcommand{\bra}[1]{\left\langle #1 \right|}
\newcommand{\rb}[1]{\left( #1 \right)}

\newcommand{\ew}[1]{\left\langle #1 \right\rangle}
\newcommand{\abs}[1]{\left| #1 \right|}
\newcommand{\beq}{\begin{eqnarray}}
\newcommand{\eeq}{\end{eqnarray}}

\newcommand{\SI}[2]{S_\mathrm{I}^{(#1,#2)}}

\newcommand{\eq}[1]{Eq.~(\ref{#1})}
\newcommand{\fig}[1]{FIG.~\ref{#1}}

\begin{document}
\title{Quantum impurity models with the Coupled Cluster Method}
\author{Jin-Jun Liang, Clive Emary, and Tobias Brandes}
\affiliation{
Institut f\"ur Theoretische Physik,
Hardenbergstr. 36,
TU Berlin,
D-10623 Berlin,
Germany
}

\date{\today}
\begin{abstract}
We investigate the ground-state properties of the Anderson single impurity model (finite Coulomb impurity repulsion) with the Coupled Cluster Method.  We consider different CCM reference states and approximation schemes and make comparison with exact Green's function results for the non-interacting model and with Brillouin-Wigner perturbation theory for the full interacting model. Our results show that coupled cluster techniques are well suited to quantum impurity problems.
\end{abstract}

\maketitle

The coupled cluster method (CCM) is a powerful method for investigating quantum many-body systems \cite{Bishop91,Bishop98}. 
Being derived from first principles, the CCM is universally applicable to many different fields, 
and its reputation, for being numerically accurate at reasonable computation costs, has been  well 
established in nuclear physics \cite{Kuemmel78,Kuemmel84}, quantum chemistry \cite{Bartlett91,Paldus99}, and quantum magnetism \cite{Bishop91a,Farnell02,ric08}. Although the method finds most application for discrete models, such as those of fields mentioned above, it has also been succesfully applied to continuum systems, such as in
Refs. \cite{Bishop78,Bishop82}. In this paper, we consider the application of the  CCM to the Anderson impurity model \cite{Anderson61}, which consists of a single localised orbital coupled to a contiuum of electronic states.

Since its inception, interest in Anderson-like impurity models has been high, as these models play important roles in strongly correlated systems \cite{Runge96,Gunnarsson83,Stewart84}, and transport through quantum dots \cite{Meir93}.
Over the years, the Anderson model has been studied with many different approachs: Bethe ansatz techniques
are used to exactly describe ground-state properties \cite{Wiegmann80,Wiegmann83,Teng1995}, 
and the Green's function method is used for an exact solution of the non-interacting 
case \cite{Gunnarsson83,Hewson93,Mahan90}.
Approximation methods, such as the variational method \cite{Varma76}, projection operators 
\cite{Fulde91,Kang95}, Hubbard operators \cite{Hubbard63,Hubbard64}, and the numerical renormalization 
group method \cite{Bulla08} are also employed to the impurity models; an approach similar to the CCM can 
as well be found in Ref.~\cite{Schonhammer78}.

We will give a short introduction to the method, followed by an overview of the Green's function exact solution to
the non-interacting Anderson model and self-consistent perturbation solution to the interacting one.
We will then present a general disscusion of possible choices of CCM reference states and correlation operators 
for the model. Several truncation schemes to these operators will be considered for both reference states and both models with
no interaction and a finite Coulomb repulsion. The results will be compared with those of exact Green's
function and self-consistent perturbation methods. All of our calculations are done in equilibrium.


\section{Coupled Cluster Method}

We begin with a brief outline of the CCM formalism; further details can be found in Refs. \cite{Bishop91,Bishop98}. Consider a general many-body system described by a Hamiltonian $H$ and exact ground-state eigenvector, $|\Psi\rangle$, such that
\begin{equation} 
H |\Psi\rangle = E_g |\Psi\rangle
\label{ccm_eq1}
.
\end{equation} 
Our system may be described in terms of a
reference state (or cyclic vector)  $\ket{\Phi}$ and a corresponding
complete set of mutually commuting multiconfigurational creation
operators $\left\{C_I^{\dagger}\right\}$.  The set
$\left\{C_I^{\dagger}\right\}$ is defined with respect to the
reference state, such that
$C_I\ket{\Phi}=0=\langle\Phi|C_I^{\dagger},~~\forall ~I\ne 0$, in
a notation in which $C_0^{\dagger} \equiv 1$, the identity
operator. In general, $I$ is a set index and the operators
$C_I^{\dagger}$ contain products of single-particle operators. The
set index $\left\{ I \right\}$ is complete in the sense that the
set of states $\left\{C^{\dagger}_I\ket{\Phi}\right\}$ provides
a complete basis for the Hilbert (or Fock) space.  The reference state,
$\ket{\Phi}$, must be chosen to be non-orthogonal to the actual
wavefunction of the system, $ \langle \Phi \ket{\Psi}\ne 0$ and thus $|\Phi\rangle$ plays the role of a vacuum 
state with respect to a suitable set of (mutually commuting) many-body 
creation operators $\{C_I^{\dagger}\}$.
For example, in describing an interacting Fermi gas, we might take the reference state to be the filled (non-interacting) Fermi sea, with the set of operators $\{C_I^{\dagger}\}$ creating all possible electron-hole excitations in this sea \cite{Bishop78,Bishop82}.

Within the single-reference CCM, the ket state of \eq{ccm_eq1} is parametrised as
\begin{eqnarray} 
|\Psi\rangle = {\rm e}^S |\Phi\rangle \; &;&  
\;\;\; S=\sum_{I \neq 0} s_I C_I^{\dagger}
\label{ccm_eq2} 
,
\end{eqnarray} 
with $S$ the CCM correlation operator and $\left\{s_I\right\}$ complex coefficients.
This exponentiated form of the ground-state CCM 
parametrisation of Eq. (\ref{ccm_eq2}) ensures the correct counting of 
the independently-excited correlated many-body clusters with respect to $|\Phi\rangle$ that are present in the exact ground state $|\Psi\rangle$. It also ensures the 
exact incorporation of the Goldstone linked-cluster theorem \cite{Goldstone57},
which itself guarantees the size-extensivity of all relevant 
extensive physical quantities. 

The eigen bra-state of our Hamiltonian is defined via the Schr\"odinger equation
\begin{eqnarray} 
\langle\tilde{\Psi}| H &=& E_g \langle\tilde{\Psi}|
.
\end{eqnarray} 
Within the normal CCM \cite{Bishop91,Bishop98}, this bra state is parameterised independently as
\begin{eqnarray} 
\langle\tilde{\Psi}| = \langle\Phi| \tilde{S} {\rm e}^{-S} \; &;& 
\;\;\; \tilde{S} =1 + \sum_{I \neq 0} \tilde{s}_I C_I.
\label{ccm_eq3}
\end{eqnarray} 
The exact groundstate eigen-bra
is given by the Hermitian adjoint of state $\ket{\Phi}$, but if the correlation operator $S$ is truncated,
the approximate eigenket $\ket{\Phi}$ may be no longer normalisable with itself and the adjoint-state ill-defined.
We note that although Hermiticity for a truncated $S$
is lost, the normalisation conditions 
$ \langle \tilde{\Psi} | \Psi\rangle
= \langle \Phi | \Psi\rangle 
= \langle \Phi | \Phi \rangle \equiv 1$ are explicitly 
imposed.

The ground-state properties of the system, then, are completely described by the set of 
CCM {\it correlation coefficients} $\{ s_I, \tilde{s}_I \}$ 
which are regarded as independent variables.
An arbitrary operator $A$ will have a ground-state expectation value given as
\begin{equation} 
\bar{A}
\equiv \langle\tilde{\Psi}\vert A \vert\Psi\rangle
=\langle\Phi | \tilde{S} {\rm e}^{-S} A {\rm e}^S | \Phi\rangle
=\bar{A}\left( \{ s_I,\tilde{s}_I \} \right). 
\label{ccm_eq6}
\end{equation} 
In particular, the ground-state energy expectation 
functional $\bar{H} ( \{ {s}_I, \tilde{{s}}_I\} )$ is given by
\begin{equation} 
\bar{H}\left( \{ s_I,\tilde{s}_I \} \right) 
\equiv \langle\tilde{\Psi}\vert H \vert\Psi\rangle
=\langle\Phi | \tilde{S} {\rm e}^{-S} H {\rm e}^S | \Phi\rangle.
\end{equation} 
By requiring  $\bar{H} ( \{ {s}_I, \tilde{{s}}_I\} )$ to be stationary with respect to variations in each of the (independent) correlation coefficients, one arrives at an expression for the ground-state energy
\begin{equation} 
E_g = E_g ( \{{s}_I\} ) = \langle\Phi| {\rm e}^{-S} H {\rm e}^S|\Phi\rangle,
\label{ccm_eq9}
\end{equation}  
and the following coupled set of equations for the coefficients
\begin{eqnarray} 
\delta{\bar{H}} / \delta{\tilde{{s}}_I} =0 & \Rightarrow &   
\langle\Phi|C_I {\rm e}^{-S} H {\rm e}^S|\Phi\rangle = 0 ,  \;\; I \neq 0 
\;\; ; \label{ccm_eq7} \\ 
\delta{\bar{H}} / \delta{{s}_I} =0 & \Rightarrow & 
\langle\Phi|\tilde{S} {\rm e}^{-S} [H,C_I^{\dagger}] {\rm e}^S|\Phi\rangle 
= 0 , \;\; I \neq 0 \; . \label{ccm_eq8}
\end{eqnarray}  
The similarity transforms may be evaluated with help of the identity
\beq
e^{-S} H e^S = H + \left[H,S\right]+\textstyle{\frac{1}{2!}}\left[\left[H,S\right],S\right] + \ldots,
\eeq
which is guaranteed to terminate since $H$ will contain only a finite number of annihilation operators $C_I$.

It is important to notice that this (bi-)variational formulation 
does not provide an upper bound for $E_g$ when the summations for 
$S$ and $\tilde{S}$ are truncated, due to the lack of 
exact Hermiticity when such approximations are made.

\section{Anderson Model}

We will study here the Anderson model of a single impurity coupled to a single reservoir. This model was originally
introduced to describe magnetic impurities in metals \cite{Anderson61}, and also finds application in describing transport through quantum dots \cite{Meir93}.  The Anderson Hamiltonian reads
\begin{align}
H= & \sum_{k,\sigma}\varepsilon_{k}c_{k\sigma}^{\dagger}c_{k\sigma}+\varepsilon_{d}\sum_{\sigma}d_{\sigma}^{\dagger}d_{\sigma}+Un_{\uparrow}n_{\downarrow}\label{H_am}\nonumber \\
 & \,+\sum_{k,\sigma}\left(V_{k}d_{\sigma}^{\dagger}c_{k\sigma}+V_{k}^{*}c_{k\sigma}^{\dagger}d_{\sigma}\right),
\end{align}
where $d^{\dagger}_\sigma$ is the creation operator of an electron of spin $\sigma$ on the localised level and $c^\dag_{k\sigma}$ is the creation operator of a continuum electron with quantum numbers $k$ and $\sigma$.  The energy of the dot level is $\varepsilon_d$, $\varepsilon_k$ is the energy of a continuum level, $U$ is the Coulomb interaction between two impurity electrons on the local level
and $V_k$ describes the coupling between dot and continuum state $k$.  For simplicity, we have assumed that $\varepsilon_d$, $\varepsilon_k$ and $V_k$ are spin-independent.

We further specify the continuum as being a band of width $2W$ with centre  chosen as our energy zero.  We chose the number of electrons in the system, $N_e$, to be that given by filling the band from $-W$ to $0$ at zero temperature.
\footnote{For calculations from Reference state II, the number of electrons included in the calculation is actually one greater than this. However, since $N$ is large, this addition has no bearing on the result}

The important quantity that describes the hybridisation of the localised level with the reservoir is the half tunnelling rate,
\begin{equation}
  \Gamma(\varepsilon)\equiv\pi g(\varepsilon)V^*_{k(\varepsilon)}V_{k(\varepsilon)}
  ,
\end{equation}
where $g(\varepsilon)$ is the density of states of the band, $g(\varepsilon)=\sum_k\delta(\varepsilon-\varepsilon_k)$.  In obtaining numerical results, we will assume that this rate is constant across the energy range of interest, $\Gamma(\varepsilon)=\Gamma$, although our technique is not limited to this approximation.

In the continuum limit and with this assumption, summations over $k$ can be converted to integrals as follows
\begin{eqnarray}
\sum_{k}\left|V_{k}\right|^{2}  &=&\nonumber
	\int_{-W}^{W}d\varepsilon\sum_k\delta(\varepsilon-\varepsilon_k)\left|V_{k}\right|^2 \\
&=&	\int_{-W}^{W}g(\varepsilon)V^2(\varepsilon)d\varepsilon =
	\frac{\Gamma}{\pi}\int_{-W}^{W} d\varepsilon. \label{sum2int}
\end{eqnarray}

\subsection{Fano-Anderson model}

Without Coulomb interaction, $U=0$, the Anderson model reduces to the Fano-Anderson, or single resonant level, model \cite{Fano61,Anderson61}.  In this case the two spin channels become  independent, and we need only consider the Hamiltonian for a single spin species:
\begin{equation} \label{H_fam}
H=\varepsilon_{d}d^{\dagger}d+\sum_{k}\varepsilon_{k}c_{k}^{\dagger}c_{k}+\sum_{k}\left(V_{k}d^{\dagger}c_{k}+V_{k}^{*}c_{k}^{\dagger}d\right)
,
\end{equation}
with omitted spin index.

The Fano-Anderson model permits an exact solution by the Green's function method  \cite{Hewson93}.  For a 
constant tunnelling rate $\Gamma$, and provided that dot level $\varepsilon_d$ lies well within the conduction band \cite{Hewson93},
the ground-state expectation value of the dot occupation number is
given by
\begin{equation}\label{ocp_grn}
\ew{n_d}(W) = \frac{\Gamma}{\pi}\int_{-W}^{0}\frac{d\omega}
{\left(\omega-\varepsilon_{d}-\Lambda(\omega)\right)^{2}+\Gamma^{2}}
,
\end{equation}
where 
$\Lambda(\omega)=\frac{\Gamma}{\pi}\ln\left|\frac{\omega+W}{\omega-W}\right|$.
In the infinite bandwidth limit, $W \to \infty$, we obtain
\begin{equation}\label{ocp_infty_grn}
\ew{n_d} = \frac{1}{2}-\frac{1}{\pi}\arctan \left(\frac{\varepsilon_d}{\Gamma}\right).
\end{equation}

\subsection{Self-consistent perturbation theory}

It will be instructive to compare our results for the Anderson model with some results already known in the literature \cite{Gunnarsson83,Mahan00}.  We will compare with 
self-consistent perturbation (SCP) results as these bear a close resemblance to the lowest-order CCM solutions.  In Ref.~\cite{Mahan00}, 
two different ansatz schemes are posited for ground state eigen-ket of the system:
\begin{eqnarray}
	\ket{\Psi^{(\mathrm{I})}} &=& \alpha_\mathrm{I} \left\{1 + \sum_{q\leqslant k_\mathrm{F},\sigma}
						\beta_q d_\sigma^{\dagger}c_{q\sigma} \right\} \ket{\mathrm{F}} 
				,\label{scpI}\\
	\ket{\Psi^{(\mathrm{II})}} &=& 
	\alpha_\mathrm{II} 
	\left\{
	1
	+ \rb{\sum_{p>k_\mathrm{F}}\eta_p c_{p\downarrow}^\dagger}  d_\downarrow
	\right.
	\nonumber\\
	&&
	\left.
	~~~~~~~~~~~~~~~
	+ \sum_{q\leqslant k_\mathrm{F}}
	\gamma_q d_\uparrow^{\dagger}c_{q\uparrow}
	\right\} 
	d_\downarrow^\dagger\ket{\mathrm{F}}
	\label{scpII}
\end{eqnarray}
where
\beq
  \ket{\mathrm{F}} \equiv
  \sum_{q\leqslant k_\textsc{f}\sigma}c^{\dagger}_{q\sigma}\ket{0}
\eeq
is the unperturbed filled Fermi sea, $\alpha_\mathrm{I,II}$ are normalization factors, and $\beta_{q}$, $\eta_p$ and $\gamma_{q}$ are variational parameters.
An expression for the ground-state energy of the Anderson model can then be obtained 
with the variational method which is here equivalent to the Brillouin-Wigner perturbation method \cite{Slivert72}. 
It is convenient to present these results in terms of 
\begin{equation} \label{dEdef}
	\delta E \equiv E_g - E_\mathrm{F}
	,
\end{equation}
the difference between the actual ground-state energy, $E_g$ and the energy of the unperturbed filled Fermi sea, $E_\mathrm{F}$.

For each ansatz, this procedure results in an equation for $\delta E $ which must be solved self-consistently.  From the first SCP ansatz, we obtain
\begin{eqnarray}\label{dE_CI1}
	\delta E   =  2\sum_{k\leq k_{F}}\frac{\left|V_{k}\right|^{2}}{\delta E+\varepsilon_{k}-\varepsilon_{d}}
	 =  2\Gamma\int_{-W}^{0}\frac{d\varepsilon}{\pi}\frac{1}{\delta E+\varepsilon-\varepsilon_{d}},\;\;
\end{eqnarray}
and from the second
\begin{eqnarray}\label{dE_CI2}
	\delta E &=& \varepsilon_{d}
		+\Gamma\int_{0}^{W}\frac{d\varepsilon}{\pi}\frac{1}{\delta E-\varepsilon}
		\nonumber\\& & \quad
		+\Gamma\int_{-W}^{0}\frac{d\varepsilon}{\pi}\frac{1}{\delta E+\varepsilon-2\varepsilon_{d}-U}
		.
\end{eqnarray}
We will discuss the nature of these solutions later, but let us note here that \eq{dE_CI1} from the first ansatz is $U$-independent, and that, since these results were obtained from a (true) variational principle, they provide upper bounds for the ground-state energy of the system.

\section{Application of CCM}
The application of the method starts with choosing an appropriate reference state $\ket{\Phi}$
and correlation operator $S$. Since for large positive $\varepsilon_d$, the impurity
level has weak correlation with the Fermi sea, one obvious choice for a 
reference state is the uncoupled Fermi sea with empty impurity level
\begin{equation}\label{ref_I}
	\ket{\Phi^{(\mathrm{I})}} = \ket{\mathrm{F}}=\sum_{q\leqslant k_\textsc{f}\sigma}c^{\dagger}_{q\sigma}\ket{0}
	.
\end{equation}
We will denote this choice as reference state I. 
With the impurity level slightly below the Fermi level, we might expect the ground-state of the system to be approximately given by a filled Fermi sea plus a singly-occupied impurity. 
We will therefore also consider an application of the CCM based on the second reference state (II)
\begin{equation}\label{ref_II}
	\ket{\Phi^{(\mathrm{II})}} = d_{\downarrow}^{\dagger}\ket{F}
	=
	d_{\downarrow}^{\dagger}\sum_{q\leqslant k_\textsc{f},\sigma}c^{\dagger}_{q\sigma}\ket{0}
	.
\end{equation}
Note that these two reference state are also the reference states for the SCP ansatz schemes in \eq{scpI} and \eq{scpII}
\footnote{The SCP ansatz schemes are then seen to be wavefunctions of the configuration interaction type with a simple $(\mathbbm{1}+F)\ket{\Phi}$ structure}.
In the following, we will consider each reference state in turn, giving the correlation operators and deriving the CCM equations for each.

\subsection{Reference state I}

The correlation operators $S$ consist of terms with operators promoting 
electrons in reference states to vacant orbitals. Assuming
particle-number conservation, the various terms of $S$ can be classified with the number of holes created in the Fermi sea, $n$, and the number of electrons created on the dot, $m$ (the remaining $n-m$ electrons are promoted to states in the continuum above the Fermi level).
For reference state I with Fermi level at $\varepsilon_\mathrm{F}=0$, we may write the complete CCM correlation operator as
\begin{eqnarray}
	S^{(\mathrm{I})} &=& \sum_{n=1}^N
	\sum_{m=0}^{\min(n,2)} S_{{\mathrm{I}}}^{(n,m)},
	\label{S_I}
\end{eqnarray}
with
\begin{eqnarray}
	S_\mathrm{I}^{(n,0)} &=& \sum_{\{pq\sigma\}_n}s_{\{pq\sigma\}_n}^{(\mathrm{I},n,0)}\prod_{j=1}^n c^{\dagger}_{p_j\sigma_j}c_{q_j\sigma_j},
	\label{S_In0}	\\
	S_\mathrm{I}^{(n,1)} &=& \sum_{\{pq\sigma\}_n}s_{\{pq\sigma\}_n}^{(\mathrm{I},n,1)}d^{\dagger}_{\sigma_1} 
				c_{q_1\sigma_1}\prod_{j=2}^n c^{\dagger}_{p_j\sigma_j}c_{q_j\sigma_j},
	\label{S_In1}	\\
	S_\mathrm{I}^{(n,2)} &=& \sum_{\{pq\sigma\}_n}s_{\{pq\sigma\}_n}^{(\mathrm{I},n,2)}d^{\dagger}_{\uparrow}c_{q_1\uparrow} 
				d^{\dagger}_{\downarrow}c_{q_2\downarrow} \prod_{j=3}^n c^{\dagger}_{p_j\sigma_j}c_{q_j\sigma_j},\quad
	\label{S_In2}
	.
\end{eqnarray}
The summations are performed over all relevant combination of indices with $p>k_{\mathrm{F}},q<k_{\mathrm{F}}$, with $k_{\mathrm{F}}$
being the Fermi wave-number.

An exact description of the interacting $U\ne0$ model would require that we keep all the above terms in the correlation operator.  This is impractical, however, and to make progress we must truncate $S$ in some fashion.

The simplest nontrivial truncation scheme is to keep only the single term,
\begin{equation}
	S_{1} = \SI{1}{1}
	.
\end{equation}
We will refer to this scheme as the $S_1$ or SUB-1 approximation, and for simplicity, relabel the relevant coefficient as
\begin{equation} \label{s1sub1}
	s_{q\sigma}^{(\mathrm{I},1,1)}= s_{q\sigma}^{(1)} 
	.
\end{equation}
With this truncation of the $S$-operator, the CCM expression for the ground-state  energy of the Anderson model reads
\begin{equation}\label{Eg_am_sub1I}
	E_{g}=\bra{\phi}e^{-S_{1}}He^{S_{1}}\ket{\phi} =2\sum_{q}\varepsilon_{q\sigma} + \sum_{q\sigma}V_{q\sigma}^{*}s_{q\sigma}^{(1)}.
\end{equation}
From \eq{ccm_eq7}, the coefficient $s_{q\sigma}^{(1)}$ is determined by the equation:
\begin{equation}\label{s1_am_sub1I}
	V_{q}+\left(\varepsilon_{d}-\varepsilon_{q}\right)s_{q\sigma}^{(1)}-s_{q}^{(1)}\sum_{q'\leq k_{F}}V_{q'}^{*}s_{q'\sigma}^{(1)}=0,
\end{equation}
which is a system of $N_e$ coupled quadratic equations.  This should be compared with 
the work in Ref. \cite{YZhou95} on the periodic Anderson model, in which $N_e$ independent quadratic equations for each parameter were described.

Under the assumption of no external magnetic field, the coupling constants $V_k$ are spin symmetric such that the correlation coefficients can be written 
\begin{equation} \label{spinsym1}
	s_{q\uparrow}^{(1)} = s_{q\downarrow}^{(1)} 
							\equiv s_{q}^{(1)}.
\end{equation}
Using these new parameters $s_{q}^{(1)}$, the ground state energy correction from \eq{Eg_am_sub1I} becomes
\begin{equation} \label{dE_sub1}
	\delta E \equiv \sum_{q\sigma}V_{q}^{*}s_{q\sigma}^{(1)} =
			2\sum_{q}V_{q}^{*}s_{q}^{(1)} .
\end{equation}
With spin symmetry, rearranging \eq{s1_am_sub1I}, we obtain an expression of $s_{q}^{(1)}$ in terms of 
$\delta E$,
\begin{equation} \label{s_dE_S1}
	s_{q}^{(1)}=\frac{V_{q}}{\delta E/2+\varepsilon_{q}-\varepsilon_{d}},
\end{equation}
Multiplying both sides of this equation by $V_{q}^{*}$ and summing over $q$, we obtain
a self-consistent equation for $\delta E$:
\begin{equation}\label{dE_am_sub1I}
	\delta E =2\int_{-W}^{0}\frac{g(\varepsilon)V^2(\varepsilon)}
			{\delta E/2+\varepsilon_{q}-\varepsilon_{d}}.
\end{equation}
For a constant $\Gamma=\Pi g(\varepsilon)V^2(\varepsilon)$,
\begin{equation} \label{dE_S1}
	 \delta E = 2\Gamma\int_{-W}^{0}\frac{d{\varepsilon}}{\pi}
		\frac{1}{\delta E/2+{\varepsilon}-{\varepsilon}_{d}}.
\end{equation}
In this form, this is like the SCP result, \eq{dE_CI1}, except that the SCP result does not have a factor one-half in the denominator on the right hand side. This factor results from double dot occupancy, as will be discussed in the results section.

Using \eq{ccm_eq8}, the bra-state coefficient $\tilde{s}_q^{(1)}$ is obtained within the SUB-1 approximation from
\begin{equation}
	V_{q}^{*}-V_{q}^{*}\sum_{q'}\tilde{s}_{q'}^{(1)}s_{q'}^{(1)}+\left(\varepsilon_{d}-\varepsilon_{q}-\frac{\delta E}{2}\right)\tilde{s}_{q}^{(1)} = 0
	.
\end{equation}
This can be solved for $\tilde{s}_{q'}^{(1)}$ by first rearranging such that we have
\begin{equation}\label{s1_S1}
	\tilde{s}_{q}^{(1)} = \frac{V_{q}^{*}\rb{1-\sum_{q'}s_{q'}^{(1)}\tilde{s}_{q'}^{(1)}}}
				{\frac{\delta E}{2}-\left(\varepsilon_{d}+\varepsilon_{q}\right)}
\end{equation}
and then substituting this equation into itself and iterating. Denoting
\begin{equation}
	\Theta_q = \frac{V_{q}^{*}}{\frac{\delta E}{2}-\left(\varepsilon_{d}+\varepsilon_{q}\right)}
	,
\end{equation}
we find
\begin{equation} \label{stld_S1}
	\tilde{s}_{q}^{(1)} = \frac{\Theta_q}{1-\sum_{q'}\Theta_{q'}s_{q'}^{(1)}}.
\end{equation}
Therefore, once \eq{dE_S1} is solved for $\delta E$, we can immediately obtain the correlation coefficients from \eq{s_dE_S1} and \eq{stld_S1}.  The impurity occupation can then be easily calculated from \eq{ccm_eq6} as
\begin{equation} \label{ocp_S1}
	\ew{n_d} = 2\sum_q s_{q}^{(1)}\tilde{s}_{q}^{(1)}
	.
\end{equation}

We will also consider a more advanced approximation based on the Scheme I reference state.  Here we choose to keep both terms in $S$ with a single hole in the electron sea and, in order to account for the effects of the Coulomb interaction on the impurity, we will also include the lowest-lying double-occupation term.  We therefore consider the correlation operator
\begin{equation} \label{S_am}
	S_2 = \SI{1}{1} + \SI{2}{2} + \SI{1}{0}
	,
\end{equation}
where $S_\mathrm{I}^{(1,1)}$ promotes electrons from the Fermi sea to the dot, $S_\mathrm{I}^{(1,0)}$ induces electron-hole correlation within the reservoir, and $\SI{2}{2} $ promotes two electrons of opposite spin to the impurity.
For notational convenience, as in \eq{s1sub1}, we relabel the CCM coefficients as
\begin{eqnarray}\label{par_labels}
	s^{(1)}_{q\sigma} = s^{(\mathrm{I},1,1)}_{q\sigma},\quad 
	s^{(a)}_{qq'\sigma\sigma'} = s^{(\mathrm{I},2,2)}_{qq'\sigma\sigma'},\quad
	s^{(b)}_{pq\sigma} = s^{(\mathrm{I},1,0)}_{pq\sigma},
\end{eqnarray}
where $\sigma\neq\sigma'$, because of the Pauli principle.  Spin symmetry means that, together with \eq{spinsym1}, the coefficients can be rewritten as
\begin{eqnarray}
	s_{qq'}^{(a)} &\equiv& 
	s_{qq'\uparrow\downarrow}^{(a)} = s_{qq'\downarrow\uparrow}^{(a)},
	\\
	s^{(b)}_{pq} &\equiv& 
		s^{(b)}_{pq\uparrow} = s^{(b)}_{pq\downarrow},
	\\
	s_{qq'}^{(a)} &=& s_{q'q}^{(a)}.
\end{eqnarray}

With this correlation operator, we obtain the same expression for the ground-state 
energy correction as before,
\begin{equation}\label{dE_am_sum}
	\delta E  = 2\sum_{q}V_{q}^{*}s_{q}^{(1)}
	.
\end{equation}
Using \eq{ccm_eq6}, we find that the CCM expression for the ground-state occupation number operator, $n_d=\sum_{\mu}d^{\dagger}_{\mu}d_{\mu}$, is
\begin{eqnarray}
	\ew{n_d} &=& \sum_{q\mu}\tilde{s}_{q\mu}^{(1)}s_{q\mu}^{(1)}+2\sum_{\substack{qq'\\ \mu\neq\nu}}\tilde{s}_{qq'\mu\nu}^{(a)}s_{qq'\mu\nu}^{(a)},
	\nonumber\\
	&=& 2\sum_{q}\tilde{s}_{q}^{(1)}s_{q}^{(1)}+4\sum_{qq'}\tilde{s}_{qq'}^{(a)}s_{qq'}^{(a)}.
	\label{ocp_am_sum}
\end{eqnarray}

Evaluating \eq{ccm_eq7} for the above reference state and truncation scheme, we find that the ket-state CCM coefficients are determined by the following equations
\begin{eqnarray}
	s_{q}^{(1)}\left(\varepsilon_{d}-\varepsilon_{q}-\sum_{q'}V_{q'}^{*}s_{q'}^{(1)}\right)+\sum_{p}V_{p}s_{pq}^{(b)}
		& & \nonumber\\
		+\sum_{q'}V_{q'}^{*}s_{qq'}^{(a)}+V_{q} &=& 0, \label{eq_sq}
	\\
	\left(2\varepsilon_{d}-\varepsilon_{q}-\varepsilon_{q'}+U-2\sum_{q''}V_{q''}^{*}s_{q''}^{(1)}\right)s_{qq'}^{(a)}
		\quad& & \nonumber\\
		+Us_{q}^{(1)}s_{q'}^{(1)}-\sum_{q''}V_{q''}^{*}\left(s_{q}^{(1)}s_{q'q''}^{(a)}+s_{q'}^{(1)}s_{q''q}^{(a)}\right)  &=& 0,
	\label{eq_sqq}
	\\
	\left(V_{p}^{*}-\sum_{q'}V_{q'}^{*}s_{pq'}^{(b)}\right)s_{q}^{(1)}+s_{pq}^{(b)}\left(\varepsilon_{p}-\varepsilon_{q}\right) &=& 0.\;
	\label{eq_spq}
	.
\end{eqnarray}
Analogously, from \eq{ccm_eq8}, the bra-state parameters are obtained from 
\begin{eqnarray} 
	V_{q}^{*}-V_{q}^{*}\sum_{q'}\tilde{s}_{q'}^{(1)}s_{q'}^{(1)}+\left(\varepsilon_{d}-\varepsilon_{q}-\frac{\delta E}{2}\right)\tilde{s}_{q}^{(1)}
		\qquad& & \nonumber\\
		-4V_{q}^{*}\sum_{q'q'',}\tilde{s}_{q'q''}^{(a)}s_{q'q''}^{(a)}
		-4\sum_{q'q''}V_{q''}^{*}\tilde{s}_{q'q}^{(a)}s_{q'q''}^{(a)} 
		& & \label{s1tld_sum_am}\\
		+4U\sum_{q'}\tilde{s}_{qq'}^{(a)}s_{q'}^{(1)}
		+\sum_{p}\tilde{s}_{pq}^{(b)}\left(V_{p}^{*}-\sum_{q'}V_{q'}^{*}s_{pq'}^{(b)}\right)
	&=& 0, \nonumber \\
	V_{q'}^{*}\tilde{s}_{q}^{(1)}+\left(2\varepsilon_{d}-\varepsilon_{q}-\varepsilon_{q'}+U-\delta E\right)\tilde{s}_{q'q}^{(a)}\qquad\quad
		& & \label{satld_sum_am}\\
	-\left(V_{q'}^{*}\sum_{q''}\tilde{s}_{qq''}^{(a)}s_{q''}^{(1)}+\sum_{q''}\tilde{s}_{q''q'}^{(a)}s_{q''}^{(1)}\right)
	&=& 0, \nonumber\\
	V_{p}\tilde{s}_{q}^{(1)}+\left(\varepsilon_{p}-\varepsilon_{q}\right)\tilde{s}_{pq}^{(b)}-V_{q}^{*}\sum_{q'}s_{q'}^{(1)}\tilde{s}_{pq'}^{(b)}.
	&=& 0.\qquad \label{sbtld_sum_am}
\end{eqnarray}

To further simplify our equations, we scale all energies with the rate $\Gamma$, and use a bar to identify scaled quantities, e.g.\ $\overline{V}_k=V_k/\Gamma $.  We then scale the CCM coefficients as follows:
\begin{eqnarray}
	s_{q(\varepsilon)}^{(1)}=\overline{V}_{q(\varepsilon)}\xi_{1}(\bar{\varepsilon}),&\, & \tilde{s}_{q(\varepsilon)}^{(1)}=\overline{V}_{q(\varepsilon)}^{*}\widetilde{\xi}_{1}(\bar{\varepsilon}), \nonumber\\
	s_{q(\varepsilon')q(\varepsilon)}^{(a)} &=& \overline{V}_{q(\varepsilon')}\overline{V}_{q(\varepsilon)}\xi_{a}(\bar{\varepsilon}',\bar{\varepsilon}), \label{scaling}\\
	\tilde{s}_{q(\varepsilon')q(\varepsilon)}^{(a)} &=& \overline{V}_{q(\varepsilon')}^{*}\overline{V}_{q(\varepsilon)}^{*}\widetilde{\xi}_{a}(\bar{\varepsilon}',\bar{\varepsilon}), \nonumber\\
	s_{p(\rho)q(\varepsilon)}^{(b)} &=& \overline{V}_{p(\rho)}^{*}\overline{V}_{q(\varepsilon)}\xi_{b}(\bar{\rho},\bar{\varepsilon}),\nonumber\\
	\tilde{s}_{p(\rho)q(\varepsilon)}^{(b)} &=& \overline{V}_{p(\rho)}\overline{V}_{q(\varepsilon)}^{*}\tilde{\xi}_{b}(\bar{\rho},\bar{\varepsilon})
	,
\end{eqnarray}
to obtain a much simplified set of expressions.  Proceeding to the integral representation (\eq{sum2int}) for the sums, the ground-state energy correction $\delta \bar{E} = \delta E / \Gamma $ reads
\begin{equation}\label{dE_int_am}
	\delta\overline{E}=\frac{2}{\pi}\int_{-\overline{W}}^{0}d\bar{\varepsilon}\xi_{1}(\bar{\varepsilon}).
\end{equation}
Similarly, the dot-occupation number (\eq{ocp_am_sum}) becomes
\begin{eqnarray}
	\ew{n_{d}}&=&2\int_{-\overline{W}}^{0}\frac{d\bar{\varepsilon}}{\pi}\xi_{1}(\bar{\varepsilon})\widetilde{\xi}_{1}(\bar{\varepsilon}) \label{nd_int_am}\\
	& &+4\int_{-\overline{W}}^{0}\frac{d\bar{\varepsilon}'}{\pi}\int_{-\overline{W}}^{0}\frac{d\bar{\varepsilon}}{\pi}\widetilde{\xi}_{a}(\bar{\varepsilon}',\bar{\varepsilon})\xi_{a}(\bar{\varepsilon}',\bar{\varepsilon}). \nonumber
\end{eqnarray}
Finally, the integral form of the CCM Scheme I equation system \eq{eq_sq} -- \eq{sbtld_sum_am} becomes
\begin{eqnarray} 
	\xi_{1}(\bar{\varepsilon})\left[\bar{\varepsilon}_{d}-\bar{\varepsilon}-\frac{\delta\overline{E}}{2}\right]+\frac{1}{\pi}\int_{0}^{\overline{W}}d\bar{\rho}'\xi_{b}(\bar{\rho}',\bar{\varepsilon}) \nonumber \\
		+\frac{1}{\pi}\int_{-\overline{W}}^{0}d\bar{\varepsilon}''\xi_{a}(\bar{\varepsilon}'',\bar{\varepsilon})+1 & = & 0,\label{xi1}\\
	\left[2\bar{\varepsilon}_{d}-\bar{\varepsilon}-\bar{\varepsilon}'+\overline{U}-\delta\overline{E}\right]\xi_{a}(\bar{\varepsilon},\bar{\varepsilon}')+\overline{U}\xi_{1}(\bar{\varepsilon})\xi_{1}(\bar{\varepsilon}') \nonumber \\
		-\frac{1}{\pi}\int_{-\overline{W}}^{0}d\bar{\varepsilon}''\left[\xi_{1}(\bar{\varepsilon})\xi_{a}(\bar{\varepsilon}',\bar{\varepsilon}'')+\xi_{1}(\bar{\varepsilon}')\xi_{a}(\bar{\varepsilon}'',\bar{\varepsilon})\right] & = & 0,\quad\;\;\\
	\xi_{1}(\bar{\varepsilon})\left[1-\frac{1}{\pi}\int_{-\overline{W}}^{0}d\bar{\varepsilon}''\xi_{b}(\bar{\rho},\bar{\varepsilon}'')\right]+(\bar{\rho}-\bar{\varepsilon})\xi_{b}(\bar{\rho},\bar{\varepsilon}) & = & 0, \label{xi2b}
\end{eqnarray}
and
\begin{eqnarray} 
	1-\overline{n}_{d}+\left(\bar{\varepsilon}_{d}-\bar{\varepsilon}-\frac{\delta E}{2}\right)\xi_{1}\left(\bar{\varepsilon}\right)
		\qquad\qquad\qquad&  &\nonumber \\
		+4\overline{U}\int_{-\overline{W}}^{0}\frac{d\bar{\varepsilon}'}{\pi}\xi_{1}\left(\bar{\varepsilon}'\right)\widetilde{\xi}_{a}(\bar{\varepsilon}',\bar{\varepsilon})
 		\qquad\qquad\qquad&  & \nonumber \\ 
		-4\int_{-\overline{W}}^{0}\frac{d\bar{\varepsilon}'}{\pi}\int_{-\overline{W}}^{0}\frac{d\bar{\varepsilon}''}{\pi}\xi_{a}(\bar{\varepsilon}'',\bar{\varepsilon}')\widetilde{\xi}_{a}(\bar{\varepsilon}',\bar{\varepsilon})
 		\quad&  & \nonumber \\ 
		+\int_{0}^{\overline{W}}\frac{d\bar{\rho}'}{\pi}\left(1-\int_{-\overline{W}}^{0}\frac{d\bar{\varepsilon}'}{\pi}\xi_{b}\left(\bar{\rho}',\bar{\varepsilon}'\right)\right)
	&=& 0, \quad\;\;\\
	\widetilde{\xi}_{1}\left(\bar{\varepsilon}_{2}\right)+\left(2\bar{\varepsilon}_{d}-\bar{\varepsilon}_{1}-\bar{\varepsilon}_{2}+\overline{U}-\delta E\right)\overline{\xi}_{a}\left(\bar{\varepsilon}_{1},\bar{\varepsilon}_{2}\right)
 		&  & \nonumber \\ 
		-\int_{-\overline{W}}^{0}\frac{d\bar{\varepsilon}'}{\pi}\xi_{1}\left(\bar{\varepsilon}'\right)\left(\overline{\xi}_{a}\left(\bar{\varepsilon}',\bar{\varepsilon}_{2}\right)+\overline{\xi}_{a}\left(\bar{\varepsilon}_{1},\bar{\varepsilon}'\right)\right)
	&=& 0, \\
	\widetilde{\xi}_{1}\left(\bar{\varepsilon}\right)+\left(\bar{\rho}-\bar{\varepsilon}\right)\widetilde{\xi}_{b}\left(\bar{\rho},\bar{\varepsilon}\right)-\int_{-\overline{W}}^{0}\frac{d\bar{\varepsilon}'}{\pi}\xi_{1}\left(\bar{\varepsilon}'\right)\widetilde{\xi}_{b}\left(\bar{\rho},\bar{\varepsilon}'\right)
	&=& 0 \label{xitld},
\end{eqnarray}
for all $\bar{\varepsilon},\bar{\varepsilon}_1,\bar{\varepsilon}_2\leqslant 0$ and $\bar{\rho}>0$.

\subsection{Reference state II}

For the second reference state, \eq{ref_II}, the correlation operator can be written as
\begin{eqnarray}
	S^{(\mathrm{II})} &=& 
	\sum_{n=1}^N 
	\rb{
          S_{{\mathrm{II}}}^{(n,1)} 
          + S_{{\mathrm{II}}}^{(n,2)}
          + S_{{\mathrm{II}}}^{(n-1,0)}
          + S_{{\mathrm{II}}}^{(n-1,\bar{1})}
        }.
	\nonumber\\
	\label{S_II}
\end{eqnarray}
with
\begin{eqnarray}
	S_{\mathrm{II}}^{(n,1)} &=& \sum_{\{pq\sigma\}_n}s_{\{pq\sigma\}_n}^{(\mathrm{II},n,1)}\prod_{j=1}^n c^{\dagger}_{p_j\sigma_j}c_{q_j\sigma_j},
	\label{S_IIn1}	\\
	S_{\mathrm{II}}^{(n,2)} &=& \sum_{\{pq\sigma\}_n}s_{\{pq\sigma\}_n}^{(\mathrm{II},n,2)}d^{\dagger}_{\uparrow} 
				c_{q_1\uparrow}\prod_{j=2}^n c^{\dagger}_{p_j\sigma_j}c_{q_j\sigma_j},
	\label{S_IIn2}	\\
	S_{\mathrm{II}}^{(n,0)} &=& \sum_{\{pq\sigma\}_n}s_{\{pq\sigma\}_n}^{(\mathrm{II},n,0)} c^\dagger_{p_1\downarrow}
				d_{\downarrow}\prod_{j=1}^n c^{\dagger}_{p_j\sigma_j}c_{q_j\sigma_j},
	\label{S_IIn0}\\
	S_{\mathrm{II}}^{(n,\bar{1})} &:=& \sum_{\{pq\sigma\}_n}s_{\{pq\sigma\}_n}^{(\mathrm{II},n,\bar{1})}d^\dagger_\uparrow d_\downarrow
					\prod_{j=1}^n c^{\dagger}_{p_j\sigma_j}c_{q_j\sigma_j},
	\label{S_IInm1}
\end{eqnarray}
where, as for scheme I, $n$ labels the number of holes created in the Fermi sea. In this expression, the first term simply creates electron-hole-pairs in the leads, the second also adds a second electron to the impurity, the third term performs a spin-flip of the impurity electron, and the fourth creates the doubly-occupied impurity state.

In this section we will consider the simple truncation scheme for the second reference state in which we truncate
\begin{equation}\label{S_II_smp}
	S = \SI{1}{1} + \SI{1}{0},
\end{equation}
such that we keep excitations of single spin-up electrons from the below
Fermi surface to the dot ($\SI{1}{1}$) and of the spin-down impurity electron from the dot to the sea above Fermi surface ($\SI{1}{0}$).
For this truncation scheme we relabel the parameters
\begin{equation}
	s^{(\mathrm{II},1,1)}_{p\sigma}= s^{(1a)}_{p\sigma}
	,\quad
	s^{(\mathrm{II},1,0)}_{q\sigma} = s^{(1a)}_{q\sigma}.
\end{equation}
With these choices, under spin symmetry, \eq{ccm_eq7} yields
\begin{eqnarray}\label{s1_am_sub1II}
	s_{p}^{(1a)} &=& \frac{V_{p}^{*}}{\sum_{p'}V_{p'}s_{p'}^{(1a)}-\varepsilon_{p}+\varepsilon_{d}},
	\label{sp_sub1II} \\
	s_{q}^{(1b)} &=& \frac{V_{q}}{\sum_{q'}V_{q'}^{*}s^{(1b)}+\varepsilon_{q}-\varepsilon_{d}-U}.
	\label{sq_sub1II}
\end{eqnarray}
Defining $\delta E_{a}\equiv\sum_{p'}V_{p'}s_{p'}^{(1a)}$ and $\delta E_{b}\equiv\sum_{q'}V_{q'}^{*}s^{(1b)}$,
rearrangement of \eq{sp_sub1II} and \eq{sq_sub1II} gives the two self-consistent equations for $\delta E_{a}$ and $\delta E_{b}$:
\begin{eqnarray*}
	\delta E_{a} 
		& = &	 \Gamma\int_{0}^{W}\frac{d\varepsilon}{\pi}\frac{1}{\delta E_{a}-\varepsilon+\varepsilon_{d}},
	\\
	\delta E_{b} &=&	\Gamma\int_{-W}^{0}\frac{d\varepsilon}{\pi}\frac{1}{\delta E_{b}+\varepsilon-\varepsilon_{d}-U}.
\end{eqnarray*}
The ground-state energy correction is in terms of these quantities:
\begin{equation}\label{dE_am_sub1II}
	\delta E=\varepsilon_{d}+\delta E_{a}+\delta E_{b}.
\end{equation}
Like the solution with SCP in \eq{dE_CI2}, this $\delta E$ involves two integral expressions. However, here the two integrals are contained within two separate 
self-consitant integral equations, whereas in the SCP expression \eq{dE_CI2},there is just a single equation for $\delta E$.


\section{Non-interacting model}

The non-interacting model (\eq{H_fam}) is bi-linear in fermionic operators and thus can be exactly described by a CCM wavefunction with a bi-linear correlation operator.  The exact correlation from reference state I for the Fano-Anderson model is therefore
\begin{equation}
	S_2 = \SI{1}{1} + \SI{1}{0}
	.
\end{equation}
In this case, the exact CCM equations are as for the full Anderson model in scheme I,  except that $U=0$, $s_{qq'}^{(a)} =0$, and the spin summation is to be suppressed.  From \eq{dE_int_am} and \eq{nd_int_am}, then, the ground-state energy difference and impurity occupancy are
\begin{eqnarray}
	\delta\overline{E}&=&\int_{-\overline{W}}^{0}\frac{d\bar{\varepsilon}}{\pi}\xi_{1}(\bar{\varepsilon});\label{dE_fam}
	\\
	\ew{n_d} &=& \int_{-\overline{W}}^{0}\frac{d\bar{\varepsilon}}{\pi}\widetilde{\xi}_{1}(\bar{\varepsilon})\xi_{1}(\bar{\varepsilon}).\label{ocp_fam}
\end{eqnarray}
Under transform \eq{sum2int} and scaling \eq{scaling}, the two systems \eq{xi1} --
\eq{xitld} take the form
\begin{eqnarray}
	\xi_{1}(\bar{\varepsilon})\left[\bar{\varepsilon}_{d}-\bar{\varepsilon}-\delta\overline{E}\right]
	+\int_{0}^{\overline{W}}\frac{d\bar{\rho}}{\pi}\xi_{b}(\bar{\rho},\bar{\varepsilon})+1 & = & 0,\label{xi1_fam}\\
	\xi_{1}(\bar{\varepsilon})\left[1-\int_{-\overline{W}}^{0}\frac{d\bar{\varepsilon}'}{\pi}\xi_{b}(\bar{\rho},\bar{\varepsilon}')\right]+(\bar{\rho}-\bar{\varepsilon})\xi_{b}(\bar{\rho},\bar{\varepsilon}) & = & 0,\label{xib_fam} \\
	1+\widetilde{\xi}_{1}(\bar{\varepsilon})(\bar{\varepsilon}_{d}-\bar{\varepsilon}-\delta\overline{E})- \ew{n_d} \qquad\qquad\qquad\;& &\nonumber\\
	+\int_{0}^{\overline{W}}\frac{d\bar{\rho}}{\pi}\widetilde{\xi}_{b}(\bar{\rho},\bar{\varepsilon})\left[1-\int_{-\overline{W}}^{0}\frac{d\bar{\varepsilon}'}{\pi}\xi_{b}(\bar{\rho},\bar{\varepsilon}')\right] & = & 0,\quad\label{xi1tld_fam}\\
	\widetilde{\xi}_{1}(\bar{\varepsilon})-\int_{-\overline{W}}^{0}\frac{d\bar{\varepsilon}'}{\pi}\xi_{1}(\bar{\varepsilon}')\widetilde{\xi}_{b}(\bar{\rho},\bar{\varepsilon}')+(\bar{\rho}-\bar{\varepsilon})\widetilde{\xi}_{b}(\bar{\rho},\bar{\varepsilon}) & = & 0,\quad\;\;\label{xibtld_fam}
\end{eqnarray} 
for all $\bar{\varepsilon}\leqslant\bar{\varepsilon}_{\textsc{f}}=0$ and $\bar{\rho}>\bar{\varepsilon}_{\textsc{f}}=0$.

The number of integral equations of this system can be reduced by expressing $\xi_b$ in terms of
$\xi_1$, and $\widetilde{\xi}_b$ in terms of $\xi_1$ and $\widetilde{\xi}_1$.
This can be done by first rearranging \eq{xib_fam} into 
\begin{equation}\label{xib_fam_eq1}
	\xi_{b}(\bar{\rho},\bar{\varepsilon})=
	\frac{\xi_{1}(\bar{\varepsilon})}{\bar{\varepsilon}-\bar{\rho}}\left[1
	-\int_{-\overline{W}}^{0}\frac{d\bar{\varepsilon}'}{\pi}\xi_{b}(\bar{\rho},\bar{\varepsilon}')\right],
\end{equation}
and substituting this back in \eq{xib_fam} again to obtain
\begin{eqnarray}\label{xib_fam_eq2}
	\xi_{b}(\bar{\rho},\bar{\varepsilon}) & = & \frac{\xi_{1}(\bar{\varepsilon})}{\bar{\varepsilon}-\bar{\rho}}\biggl[1-\int_{-\overline{W}}^{0}\frac{d\bar{\varepsilon}'}{\pi}\frac{\xi_{1}(\bar{\varepsilon}')}{\bar{\varepsilon}'-\bar{\rho}} 
	\nonumber\\
	& &\qquad\qquad
	\times\left(1-\int_{-\overline{W}}^{0}\frac{d\bar{\varepsilon}''}{\pi}\xi_{b}(\bar{\rho},\bar{\varepsilon}'')\right)\biggr]; \;
\end{eqnarray}
by repeating these substitutions and rearrangements for infinitely many times, we arrive at 
\begin{equation}\label{xib_chi_fam}
	\xi_{b}(\bar{\rho},\bar{\varepsilon}) = 
	   \frac{\xi_{1}(\bar{\varepsilon})}{\bar{\varepsilon}-\bar{\rho}}\frac{1}{1+\chi(\bar{\rho})},
\end{equation}
where
\begin{equation}\label{chi_fam}
	\chi(\bar{\rho}):=\chi\left[\xi_{1}\right]\left(\bar{\rho}\right):=\int_{-\overline{W}}^{0}\frac{d\bar{\varepsilon}'}{\pi}\frac{\xi_{1}(\bar{\varepsilon}')}{\bar{\varepsilon}'-\bar{\rho}}.
\end{equation}
Similarly,
\begin{equation}\label{xibtld_chi_fam}
	\widetilde{\xi}_{b}(\bar{\rho},\bar{\varepsilon}) =
	   \frac{1}{\bar{\varepsilon}-\bar{\rho}} 
	   \left(\widetilde{\xi}_{1}(\bar{\varepsilon})-\frac{1}{1+\chi(\bar{\rho})}
	   \int_{-\overline{W}}^{0}\frac{d\bar{\varepsilon}'}{\pi}\frac{\xi_{1}(\bar{\varepsilon}')}
	   {\bar{\varepsilon}'-\bar{\rho}}\widetilde{\xi}_{1}(\bar{\varepsilon}')\right),\;
\end{equation}
Finally, with \eq{xib_chi_fam} and \eq{xibtld_chi_fam}, system \eq{xi1_fam} -- \eq{xibtld_fam} is reduced to
\begin{eqnarray} 
	\xi_{1}(\bar{\varepsilon})\biggl[\bar{\varepsilon}_{d}-\bar{\varepsilon}-\delta\overline{E}\qquad\qquad\qquad\qquad\qquad\qquad & & \label{inteq1_fam} \\ 
	+\int_{0}^{\overline{W}}\frac{d\bar{\rho}}{\pi}\frac{1}{\left(\bar{\varepsilon}-\bar{\rho}\right)\left(1+\chi\left[\xi_{1}\right]\left(\bar{\rho}\right)\right)}\biggr]+1 &=& 0,\nonumber
	\\
	1+\widetilde{\xi}_{1}(\bar{\varepsilon})(\bar{\varepsilon}_{d}-\bar{\varepsilon}-\delta\overline{E})-\int_{-\overline{W}}^{0}\frac{d\bar{\varepsilon}'}{\pi}\xi_{1}(\bar{\varepsilon}')\widetilde{\xi}_{1}(\bar{\varepsilon}') & & \nonumber\\
	+\int_{0}^{\overline{W}}\frac{d\bar{\rho}}{\pi}\frac{1}{1+\chi(\bar{\rho})}\frac{1}{\bar{\varepsilon}-\bar{\rho}} \;\qquad\qquad\qquad\qquad & & \label{inteq2_fam} \\
	\times\left(\widetilde{\xi}_{1}(\bar{\varepsilon})-\frac{1}{1+\chi(\bar{\rho})}\int_{-\overline{W}}^{0}\frac{d\bar{\varepsilon}'}{\pi}\frac{\xi_{1}(\bar{\varepsilon}')}{\bar{\varepsilon}'-\bar{\rho}}\widetilde{\xi}_{1}(\bar{\varepsilon}')\right) & = & 0.\nonumber
\end{eqnarray}

Equation (\ref{inteq2_fam}) is a linear integral equation for $\widetilde{\xi}_1$, so once 
$\xi_1$ is known, $\widetilde{\xi}_1$ can be obtained directly.  
The equation for $\xi_1$ (\eq{inteq1_fam}), however, 
is non-linear and must be solved using a numerical scheme for solving non-linear integral equations.  We use the $N$th order Legendre-Gauss Quadrature rule \cite{NRC} to discretize integrals in \eq{inteq1_fam}, thus reducing the problem to that of solving $N$ coupled non-linear algebraic equations. As exact results are available for the Fano-Anderson model, we are able to compare with these results and determine the accuracy of our numerical procedure.

\begin{figure}[t]
  \begin{center}
  \psfrag{ed}{$\varepsilon_{d}/\Gamma$} 
  \psfrag{nd}{$\ew{n_{d}}$}
  \psfrag{lndnd}{$\ln\delta n_{d}^{(N)}$}
  \psfrag{ed=0}{$\varepsilon_d/\Gamma=0$}
  \psfrag{lnN}{$\ln N$}
  \psfrag{W=50}{$\,W/\Gamma=50$}
  \psfrag{W/G=50}{$W/\Gamma=50$}
  \psfrag{ed=m2MMi}{$\varepsilon_d/\Gamma=-2$}
  \epsfig{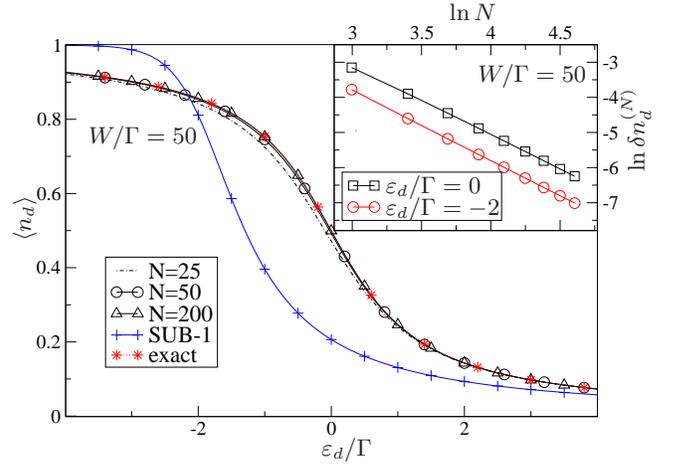}
  \caption{
	 {\bf Main panel:}
	 Ground-state impurity occupation number of the Fano-Anderson model as a function of the dot level position $\varepsilon_d$
	 with Fermi level at $\varepsilon_\mathrm{F}=0$. 
	 Plotted are the exact Green's function result (\eq{ocp_grn}) and results for the CCM $S_1$- and $S_2$-correlation operators.  
	The $S_2$ results of \eq{ocp_fam} are labelled with $N$, the number of nodes used in discretizing 
	equation system	\eq{inteq1_fam} -- \eq{inteq2_fam}.
	 The SUB-1 curve ($S_1$ result) is the analytic solution of \eq{n_S1_fam}.
	 {\bf Inset:} Investigation of numerical convergence of CCM-$S_2$ scheme.  Here we plot $\delta n_{d}^{(N)}$, which is 
	the absolute error between
	$\ew{n_d}^N$ and the exact $\ew{n_d}^{(\mathrm{exact})}$, $\delta n_{d}^{(N)}=\left|\ew{n_d}^{(\mathrm{exact})}-\ew{n_d}^N\right|$,
	as a function of $N$. 
	Results are shown for two impurity-level positions:
	 $\varepsilon_d/\Gamma=0$ and $-2$, with $N$ in the range $20$ --- $100$.  These results show the good convergence of the CCM $S_2$ scheme to the exact result.
    \label{ocpfam}
   }
  \end{center}
\end{figure}

Figure \ref{ocpfam} plots the ground-state impurity occupancy $\ew{n_d}$ of the Fano-Anderson model against dot-level position 
$\varepsilon_d$ using both $S_1$ and $S_2$ CCM
correlation operators. Also plotted is the exact result of \eq{ocp_grn}. 

The occupancy in the $S_1$-approximation is given by \eq{ocp_S1} but with no factor $2$ in front:
\begin{equation} \label{n_S1_fam}
  \ew{n_d} = \sum_q s_{q}^{(1)}\tilde{s}_{q}^{(1)}
  ,
\end{equation}
with $\delta E$ and the correlation coefficients given by their previous expressions (e.g. \eq{dE_S1} for $\delta E$) but with denominators $\delta E/2+{\varepsilon}-{\varepsilon}_{d}$ replaced by $\delta E+{\varepsilon}-{\varepsilon}_{d}$ to account for the fact that we only have one spin species here.
The $S_2$-solutions are the numerical solutions of the integral equations \eq{inteq1_fam} -- \eq{inteq2_fam}, with different numbers of discretization nodes, $N$, in the
range $-4<\varepsilon_d/\Gamma<4$.
These numerical solutions are denoted $\ew{n_d}^{(N)}$.
In the inset of this figure, we plot details of the convergence of the CCM $S_2$ calculation to the exact result as a function of $N$ ($N=20,30,\cdots,100$).
We plot results for $\varepsilon_d/\Gamma=0,-2$; CCM results for values of $\varepsilon_d$ away from zero always show less error than at  $\varepsilon_d=0$. In this inset, we plot the absolute difference $\delta n^{(N)}_d$ between the numerical results and the exact solution,
\begin{equation}
	\delta n^{(N)}_d = \abs{\ew{n_d}^{(N)}-\ew{n_d}^{(\mathrm{exact})}}
	.
\end{equation}

These results illustrate 
convergence of the numerical results of \eq{ocp_fam} to the exact solution,
as the number of discrete nodes increases. From the inset, we can determine that for the worst case at $\varepsilon_d$ around $0$, the absolute error is about $0.5\%$ for $N=100$, and the relation between the absolute error and $N$
is approximately $\mathrm{error}\propto 1/N^2$. 
This affirms the equivalence between \eq{ocp_fam} and \eq{ocp_grn}, and hence the fact that CCM is able to well reproduce the exact dot-occupation number for the Fano-Anderson model.
In the main figure, the SUB-1 curve is significantly different from the others, showing the important role of the particle-hole
correlation $\SI{1}{0}$, without which the occupation number becomes unsymmetric around the origin, and the dot level becomes narrower.


\section{Results for full Anderson model}

The presence of a finite-$U$ interaction to the full Anderson Hamiltonian, makes it impossible to be solved with a small number of correlation terms in $S$, as was the case for the Fano-Anderson model.  
In this section, we present our approximate CCM results for the ground-state energy difference of the full Anderson model, and make the comparison with results of the SCP method. 

For our CCM calculation from reference state I we will keep the double occupation term $\SI{2}{2}$ in the correlation operator but drop $\SI{1}{0}$ for simplicity. The correlation operator therefore reads
\begin{equation}\label{S_2a}
	S_{2a} = \SI{1}{1} + \SI{2}{2}.
\end{equation}	

\begin{figure}[ht]
  \begin{center}
  \psfrag{ed}{$\varepsilon_{d}/\Gamma$}
  \psfrag{dE}{$\delta E/\Gamma$}
  \psfrag{W/G=60}{$W/\Gamma=60$}
  \psfrag{U/G=6}{$U/\Gamma=6$}
  \epsfig{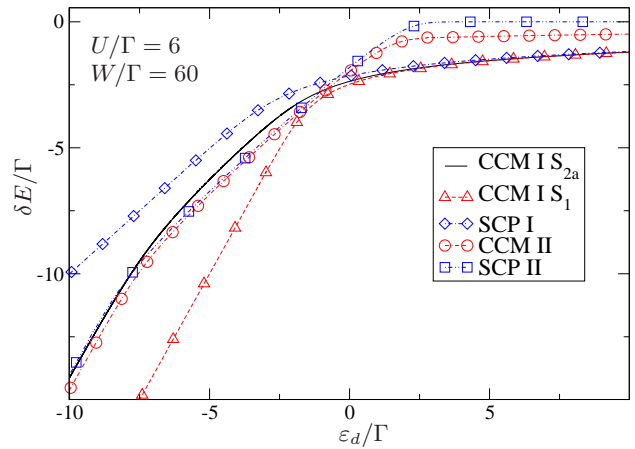}
  \caption{
	The groundstate energy difference of the Anderson model with $U/\Gamma=6$, $W/\Gamma=60$ Ref.~\cite{Herbst87_evol}. 
	The solid line is numerical result of the CCM with reference State I and correlation operator $S_{2a}$.  $N=41$ discretization nodes were used, with an estimated numerical error within $\lesssim1\%$.
	The dashed line with triangles and circles are the CCM $S_1$ solution of \eq{dE_S1} (reference state I) and and CCM II \eq{dE_am_sub1II}
	(reference states II). The remaining curves are results from self-consistent
	perturbation method, with both ansatz I \eq{dE_CI1} (diamonds) and II \eq{dE_CI2} (squares).
	The occupation of the dot at any given values of $\epsilon_d$ is roughly given by the slope of $\delta E$ curve at that point. One therefore expects that the exact result has slope $\sim 0$ in region $\varepsilon_d>0$, $1$ in $-U<\varepsilon_d<0$, and $2$ for $\varepsilon_d<-U$. Solutions CCM I$S_{2a}$, CCM II and SCP II are the only approximate solutions which capture this physics adequately.
	In addition, the SCP I and SCP II results provide exact upper bounds for $\delta E$ and we see that CCM I $S_{2a}$ curve follows this bound closely thoughout the range and thus is seen to be the best result in both aspects, in all regions.
    \label{figdE}
   }
  \end{center}
\end{figure}

Figure \ref{figdE} shows the ground-state energy correction $\delta E$, defined in \eq{dEdef} as calculated from the various methods under consideration here.
The results CCM I $S_1$, SCP I, CCM II, and SCP II are all analytic solutions:
CCM $S_1$ is the self-consistent solution of equation \eq{dE_S1}; SCP I and SCP II are solutions of \eq{dE_CI1} and \eq{dE_CI2} respectively; and CCM II is solution of \eq{dE_am_sub1II}.
The result CCM I $S_{2a}$, on the other hand, is a numerical solution of equation \eq{dE_int_am} with a finite number of discretization points $N=41$.  
From the Fano-Anderson calculation, we expect this solution to have an estimated error of $1\%$ of $\Gamma$. 
The reason why fewer discretization steps are used here as compared with the Fano-Anderson model is that for the Anderson model, the coefficient $s^{(a)}$ has two energy indices, whereas the final equation for the Fano-Anderson model only involves single-index.

The results of \fig{figdE} can be understood physically as follows.
For a vanishingly small coupling between reservoir and impurity ($\Gamma \to 0$), we can expect $\delta E$ to consist of three straight lines: for $\varepsilon>0$, we expect there to be no electron on the dot such that $\delta E\sim0$ with slope zero; for $-U<\varepsilon_d<0$, one electron from the Fermi sea will fill the impurity, yielding $\delta E \sim \epsilon_d$ (slope $1$); and finally, for $\varepsilon_d<-U$, two electrons with opposite spins will overcome Coulomb 
interaction and occupy the impurity, such that $\delta E \sim 2 \epsilon_d + U$ (slope $2$). For finite $\Gamma$, we expect the exact solution of $\delta E$ to broadly exhibit the above features, with transitions between them over an energy scale of order $\sim\Gamma$.
In the figure, CCM I $S_{2a}$, CCM II, and SCP II curves all demonstrate three regions where the solution has gradients $0$, $1$, and $2$. In contrast, the CCM I $S_1$ result only has 0 and 2 gradients, whereas SCP I has only 0 and 1 gradients. 
This difference in the negative $\varepsilon_d$ regime is a manifestation of the
exponential structure of a CCM ground state ansatz --- although there is no double occupation correlation in $S_1$, having $S_1^2$ in its expansion $e^{S_1}$, it can still produce a doubly occupied dot.

Moreover, since SCP method is equivalent to a variational calculation, the SCP results  give upper bounds for the exact value of $\delta E$. 
In the region $\varepsilon_d>0$, SCP I gives a
lowest upper bound, whereas for $\varepsilon_d<0$, it is SCP II that gives the lower one --- neither solution provides a consistent upper bound for the
whole range of $\epsilon_d$.
However, as is clear from \eq{figdE} the CCM II solution follows rather closely the best upper bound from both SCP calculations across the complete range.  This distinguishes it as, of all the approximate methods discussed here, as producing the best result across the whole parameter space of the model. Note however, that the CCM solution does not provide an upper bound.

\section{Conclusions}

Our calculations demonstrate the applicability of the CCM to quantum impurity models.  Comparisons with exact results for the non-interacting Fano-Anderson model demonstrate that the CCM can determine ground-state properties for such models to a high degree of accuracy.

In typical applications, the (truncated) CCM normally works well when the exact ground-state of the system resembles the reference state to some extend. Here, however, the method is able to produce good results for the (interacting) Anderson
model in all regimes;  even the simplest truncation scheme is able to describe double occupancy of the impurity.

In our calculations, the main technical difficulty is with solving integral equations, while higher-order CCM coeffcients will have more wave-number indices ($p$ and $q$). If we have $N$ discretization notes for each integral, $n$ indices require $N^n$ data points per coefficient. This problem can be reduced by resummation of electron-hole contributions as is done here for the Fano-Anderson model.

Future work includes the extension of our calcualtions to multiple impurity problems, such as the periodic Anderson model \cite{YZhou95}, and to include higher-order terms in the correlation operator to fully cover the Kondo effect.

\acknowledgments
This work was supported by DFG Grant BR 1528/5.

\bibliography{amccm}{}

\end{document}